\documentclass[twoside,twocolumn,tightenlines, aps,prb]{revtex4}
\usepackage{graphicx}
\usepackage{wasysym}
\usepackage{amssymb}
\usepackage[makeroom]{cancel}
\newcommand{\be}{\begin{equation}}
\newcommand{\ee}{\end{equation}}
\usepackage[unicode=true,pdfusetitle,
 bookmarks=true,bookmarksnumbered=false,bookmarksopen=false,
 breaklinks=false,pdfborder={0 0 1},backref=false,colorlinks=false]
 {hyperref}
\hypersetup{
 colorlinks,linkcolor=blue,citecolor=blue,urlcolor=blue}
 
\begin{document}

\title{Effect of proximity-induced spin-orbit coupling in graphene mesoscopic billiards}

\author{Anderson L. R. Barbosa$^1$, Jorge Gabriel G. S. Ramos$^2$, Aires Ferreira$^3$}

\affiliation{$^1$Departamento de F\'{\i}sica,Universidade Federal Rural de Pernambuco, 52171-900 Recife, Pernambuco, Brazil\\$^2$Departamento de F\'isica, Universidade Federal da Para\'iba, 58051-970 Joa\~ao Pessoa, Para\'iba, Brazil\\
$^3$Department of Physics and York Centre for Quantum Technologies, University of York, York YO10 5DD, United Kingdom}

\begin{abstract}

van der Waals heterostructures based on two-dimensional materials have recently become a very active topic of research in spintronics, both aiming at a fundamental description of spin dephasing processes in nanostructures and as a potential element in spin-based information processing schemes. Here, we theoretically investigate the magnetoconductance of mesoscopic devices built from graphene proximity-coupled to a high spin-orbit coupling material. Through numerically exact tight-binding simulations, we show that the interfacial breaking of inversion symmetry generates robust weak antilocalization even when the $z\rightarrow -z$ symmetric spin-orbit coupling in the quantum dot dominates over the Bychkov--Rashba interaction. Our findings are interpreted in the light of random matrix theory, which links the observed behavior of quantum interference corrections to a transition from a circular-orthogonal to circular-symplectic ensemble.

\end{abstract}

\maketitle

Atomically thin heterostructures of graphene and other van der Waals crystals are ideally suited for investigations of relativistic spin-orbit coupling (SOC) phenomena owing to their hybridized Dirac-like electronic structure and strong interplay between spin and lattice-pseudospin degrees of freedom [\onlinecite{RevModPhys.92.021003,C7CS00864C,PhysRevLett.121.126802,natphys.11.857}]. This thrust of research has been focused on the demonstration of $\mathbb{Z}_2$ topological insulating phases [\onlinecite{PhysRevLett.95.226801,PhysRevLett.95.146802,PhysRevB.98.081407,Wu76, PhysRevLett.120.156402,PhysRevLett.122.046403,IsingSOC}] and gate-tunable spin-charge conversion effects in graphene with proximity-induced SOC [\onlinecite{PhysRevLett.119.196801,acs.nanolett.8b04368,acs.nanolett.9b01611,natmat.19.170,acsnano.0c01037,natcomm.11.3657,PhysRevLett.124.236803}].

Interface-induced SOC in a graphene flake is detectable by low-field magnetotransport measurements, which provide a sensitive probe of symmetry-breaking processes affecting the interference between time-reversed paths of electrons [\onlinecite{PhysRevX.6.041020,Yang_2016,Yang2017,Volkl2017,Zihlmann2018,Wakamura2018}]. 
The strength of relativistic SOC effects can be estimated from the low-field changes in the magnetoconductance generated by the decrease in electron backscattering probability below its semiclassical value, a phenomenon known as weak anti-localization (WAL) [\onlinecite{PTP.63.707,PhysRevB.53.3912,PhysRevLett.108.166606,PhysRevLett.121.087701,PhysRevLett.122.156601,PhysRevB.99.205407}]. For stacked honeycomb layers of graphene and transition metal dichalcogenides with trigonal prismatic phase coordination (known as the 1H phase), simple group theory considerations show that the allowed spin-orbit interactions  comprise (spin-flip) Bychkov--Rashba (BR) interaction in addition to intrinsic-type (spin-preserving) SOC, reflecting the lowering of the point group symmetry e.g., from $D_{6h}$ to $C_{3v}$ in graphene placed on semiconducting group VI dichalcogenides  [\onlinecite{10.1038/ncomms9339, PhysRevB.93.155104,PhysRevB.95.165415,PhysRevB.98.045407,note1}]. If the  spin-orbit scattering time is short compared with the phase coherence time $\tau_{\text{so}} \ll \tau_{\phi}$, the rotation of the electron's spin in the SOC field 
generates destructive interference, and WAL [rather than conventional weak localization (WL)] is observed. The WAL magnetoconductance peaks detected in graphene/group-VI dichalcogenide bilayers at low temperatures of a few kelvins have been attributed to proximity-induced SOC in the range 1-10 meV [\onlinecite{PhysRevX.6.041020}], i.e. several orders of magnitude larger than graphene's weak intrinsic SOC ($\lambda_{\text{KM}} \approx 42 \, \mu$eV [\onlinecite{PhysRevLett.122.046403}]). Such a remarkable enhancement of SOC effects (predicted by density functional theory calculations in Refs.~[\onlinecite{10.1038/ncomms9339,PhysRevB.93.155104}]) is consistent with the strong reduction of spin lifetimes detected in Hanle-type spin precession measurements in such systems [\onlinecite{SRTA-18-Benitez, SRTA-Xu, SRTA-18-Leutenantsmeyer, SRTA-19-Omar}] and has enabled the unambiguous detection of inverse spin galvanic and spin Hall effects in graphene-based heterostructure at room temperature [\onlinecite{PhysRevLett.119.196801,acs.nanolett.8b04368,acs.nanolett.9b01611,natmat.19.170,acsnano.0c01037,natcomm.11.3657,PhysRevLett.124.236803}].

\begin{figure}[!]
\includegraphics[width=0.9\linewidth]{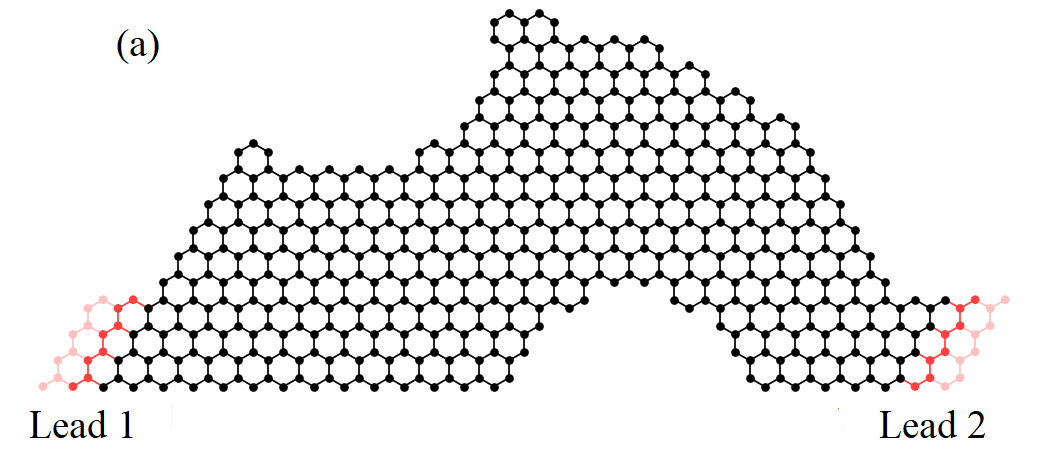}
\includegraphics[width=0.8\linewidth]{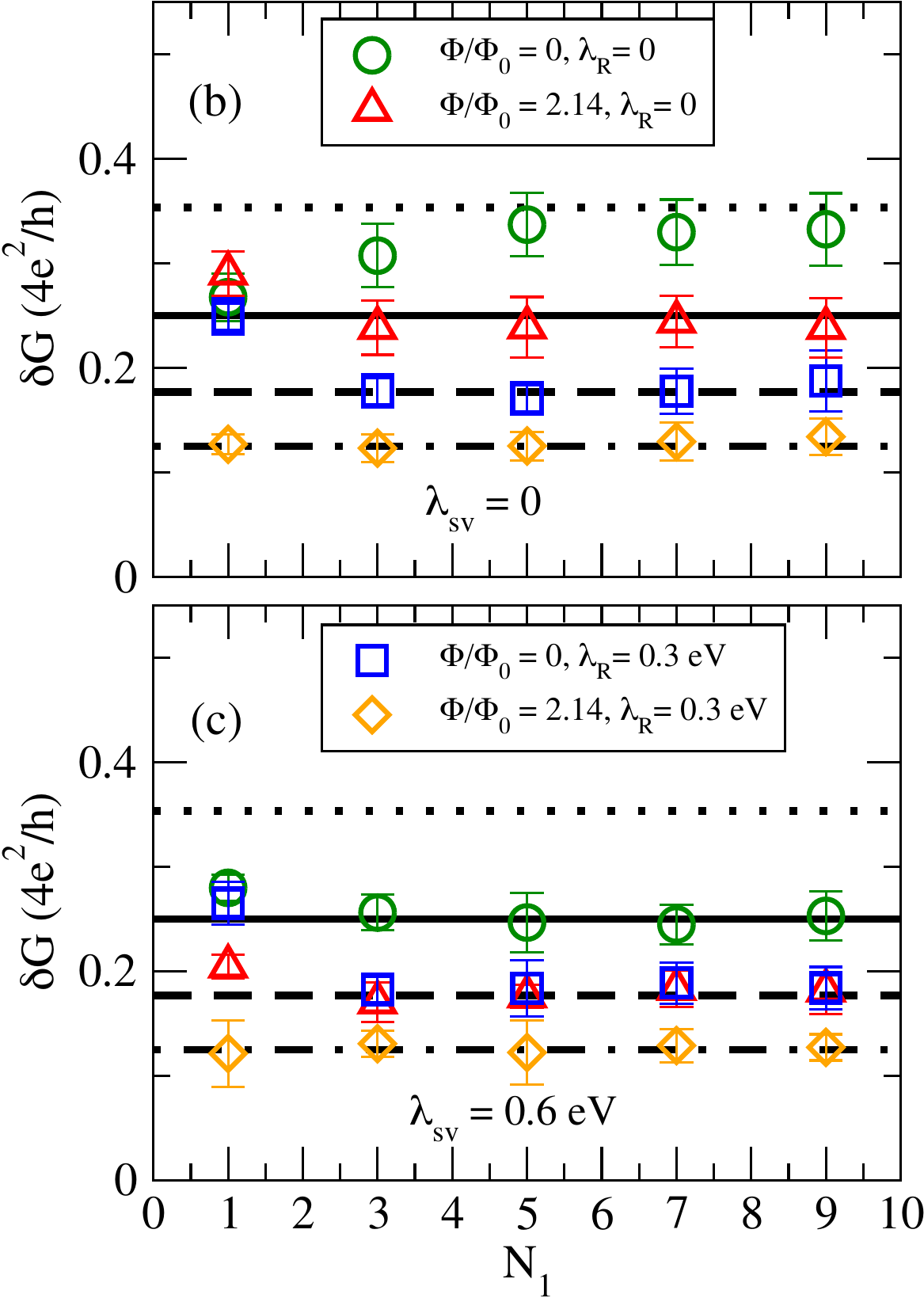}
\caption{(a) Dirac-Rashba billiard connected to two leads. (b,c) Conductance fluctuation as a function of the number of open channels for selected values of SOC and magnetic flux. Numerical data are seen to quickly converge to the predicted values in the different asymptotic limits (see Table \ref{tabela}). Here, $\Phi_0=h/e$ is the quantum of flux. See main text for other simulation parameters.}
\label{Imagem1}
\end{figure}

In this paper, inspired by this success in nanofabrication and measurement techniques, we investigate quantum coherent transport in two-dimensional open  cavities with symmetry-breaking SOC. In the envisaged Dirac quantum dot made from a van der Waals metamaterial with proximity-induced SOC (Fig.\,\ref{Imagem1}a), scattering by the irregular boundaries destroys   all constants of motion and the system is expected to exhibit   universal features of \textit{quantum chaos} [\onlinecite{RevModPhys.69.731,RevModPhys.81.539,Beenakker,RevModPhys.82.2845,PhysRevLett.87.256801,PhysRevB.65.081302,PhysRevB.99.195131,PhysRevB.84.035453}]. A unique fingerprint of quantum chaos is the emergence of universal conductance fluctuations (UCF), which are insensitive to the system size and only depend on the symmetries of the random ensembles which describe the chaotic cavity [\onlinecite{PhysRevLett.65.2442,PhysRevB.55.1142,PhysRevB.68.125329,ChoeChang,PhysRevB.93.115120,PhysRevB.93.125136}]. To date, theoretical studies of graphene-based billiards have employed Dirac models with a mass term representing a sublattice-staggered on-site energy [\onlinecite{PhysRevLett.102.056806,PhysRevB.84.205421,PhysRevB.79.161409,PhysRevB.79.115423}]. {These early studies revealed that graphene billiards behave essentially as two copies of a ``neutrino billiard'' mutually coupled through intervalley scattering  [\onlinecite{PhysRevLett.102.056806}]. If the intervalley scattering time is much shorter than the Heisenberg time required to resolve individual energy levels, the coupling between valleys restores time-reversal symmetry and the energy-level statistics are expected to follow the orthogonal symmetry class. Because the amount of valley mixing in a Dirac system can be adjusted by controlling the edge type [\onlinecite{PhysRevB.84.205421}], intervalley coupling provides a knob to tune the symmetry class.} Here, we use random matrix theory to derive a quantitative description of how proximity-induced SOC affects the statistical behavior of the conductance and its fluctuations. Our results, which extend previous studies of graphene-based chaotic billiards without SOC [\onlinecite{PhysRevLett.102.056806,PhysRevB.93.125136,PhysRevB.84.205421}], are confirmed by explicit tight-binding simulations.   

We consider a quantum dot weakly coupled to two electron reservoirs via ballistic point contacts, labeled 1 and 2, with $N_{1(2)}$ channels each (Fig.~\ref{Imagem1}a). The quantum dot is described by the scattering matrix, which for Dirac quasiparticles is a $4N \times 4N$ unitary matrix with $N=N_1+N_2$, written as

\begin{eqnarray}
\mathcal{S}=
\left[\begin{array}{cc}
r&t' \\
t &  r'
\end{array}\right], \label{Smatrix}
\end{eqnarray} 
where $t(t')$ and $r(r')$ are sub-matrices that describe transmission and reflection of Dirac quasiparticles (with spin $\sigma=\,\uparrow,\downarrow$ and sublattice pseudospin $\Sigma=\,A,B$) arriving from the lead 1 (2). Because               of the presence of two distinct spin-flavors at low energies, the $\mathcal S$ matrix in Eq.~(\ref{Smatrix}) is four times as large as that for conventional semiconductor quantum dots [\onlinecite{PhysRevB.65.081302,PhysRevB.68.125329}]. The device conductance at zero temperature is proportional to the transmission probability from lead 1 to lead 2 according to the Landauer-B\"{u}ttiker formula $G = \frac{e^2}{h}\text{Tr}\left(tt^\dagger\right)$. {In what follows, we assume that mean dwell time is much longer than the time needed for ergodic exploration of the phase space $\tau_{d} \gg \tau_{\text{erg}}\sim L/v$, so that the electron motion within the cavity is chaotic (here, $L$ is the typical cavity lateral dimension and $v$ is the Fermi velocity).\,We also assume that the intervalley scattering time is short compared to the dwell time (due to strong intervalley scattering from armchair-type edges  [\onlinecite{PhysRevB.84.205421}]), so that the conductance statistics of the open cavity are governed by the orthogonal ensemble in the absence of SOC and magnetic fields.} In the universal regime of quantum transport $\tau_{\phi}, \tau_{\text{so}} \gg \tau_{\text{erg}}$, random matrix theory can be employed to calculate the conductance distribution [\onlinecite{RevModPhys.69.731}]. In terms of the scattering matrix, the conductance reads as
\begin{equation}
G = \frac{4e^2}{h}\frac{N_1N_2}{N} -  \frac{e^2}{h} \text{Tr} \left[\mathcal{S}\, \mathcal{K}\, \mathcal{S}^{\dagger}\, \mathcal{K}\right], \label{G}
\end{equation}
where $\mathcal{K}_{i,i}=N_2/N$ ($i=1,\dots,4N_1$) and $\mathcal{K}_{i,i}=-N_1/N$ ($i=4N_1+1,\dots,4N$), and  $\mathcal{K}_{ij}=0$ for $i\neq j$. 
In order to find the average and variance of the conductance it suffices to compute the average 
\begin{eqnarray}
\left< \mathcal{S}_{ij;\alpha \beta}(\epsilon,\vec \mathcal{B}) \, \mathcal{S}_{i'j';\alpha' \beta'}^*(\epsilon,\vec \mathcal{B})\right> \label{SS}
\end{eqnarray}
in the presence of SOC and for arbitrary values of Fermi energy $\epsilon$ and magnetic field $\vec \mathcal{B}$ [\onlinecite{PhysRevB.65.081302}]. (Here, Greek indices agglutinate spin and pseudospin labels.) In what follows, we assume symmetric leads ($N_1=N_2=N/2$). The statistical properties of the scattering matrix $\mathcal{S}$ are calculated from random unitary matrices within the so-called "stub model"; see the supplemental material (SM) [\onlinecite{SM}]. After a tedious but straightforward computation, the average and variance of the dimensionless conductance per sublattice and spin ($g\equiv h G/4e^2$) can be written as follows

\begin{table*}
\centering
\begin{tabular}{cccccccc}
\hline
    \hline
Limit \quad & \quad TRS \quad & \quad SRS \quad & \quad Strong SOC \quad\quad & \quad Ensemble \quad\quad & \quad\quad $\langle g_{qc} \rangle$ \quad \quad & \quad\quad $\delta g$ \quad \quad & \quad $\frac{\delta g_{\text{COE}}}{\delta g}$ \\\hline
    \hline
$\tau_{\mathcal{B}}, \tau_{\text{sym}},\tau_{\text{asy}}\gg \tau_d$ (\text{Refs.} [\onlinecite{PhysRevB.84.205421,PhysRevB.93.125136}]) & $\checkmark$ & $\checkmark$ & No & COE & $- 0.250$ & 0.353 & 1  \\\hline
$\tau_{\text{sym}},\tau_{\text{asy}}\gg \tau_d \gg \tau_{\mathcal{B}}$ (\text{Refs.} [\onlinecite{PhysRevB.84.205421,PhysRevB.93.125136}]) & $\times$ & $\checkmark$ & No &   CUE &0.000 & 0.250 & $\sqrt{2}$\\
$\tau_{\mathcal{B}}, \tau_{\text{asy}}\gg \tau_d \gg \tau_{\text{sym}}$ & $\times$ & $\times$  & Yes & CUE &0.000 &0.250  &$\sqrt{2}$ \\\hline
$\tau_{\mathcal{B}}\gg \tau_d \gg \tau_{\text{asy}}$& $\checkmark$ & $\times$ & Yes & CSE &0.125 & 0.176 & 2 \\\hline
$\tau_{\text{asy}}\gg \tau_d \gg \tau_{\mathcal{B}}, \tau_{\text{sym}}$ & $\times$ & $\times$ & Yes & &0.000 & 0.176 & 2  \\
$\tau_d \gg \tau_{\mathcal{B}}, \tau_{\text{asy}}$ & $\times$ & $\times$ & Yes & &0.000 & 0.125 & $2\sqrt{2}$ \\\hline
    \hline
\end{tabular}
\caption{Ensemble transitions facilitated by strong SOC in chaotic Dirac-Rashba billiards according to random matrix theory [Eqs.(\ref{wl})-(\ref{var})]. COE, CUE and CSE refer to the standard circular ensembles (orthogonal, unitary and symplectic, respectively). The $\checkmark$ ($\times$) symbol indicates the presence (absence) of a given symmetry. SRS, spin rotational symmetry; TRS, time-reversal symmetry.}
\label{tabela}
\end{table*}
\begin{eqnarray}
\langle g \rangle  &=& \frac{N}{4} +  \frac{1}{8} \left[\frac{1}{1+\Gamma_{\mathcal{B}}}-\frac{1}{1+\Gamma_{\mathcal{B}}+2\Gamma_{\text{asy}}}\right.\nonumber\\ &-& \left. \frac{2}{1+\Gamma_{\mathcal{B}}+\Gamma_{\text{asy}}+\Gamma_{\text{sym}}}\right]  , \label{wl} \\
\text{var}[g] &=&\frac{1}{64}\sum_{j=0}^1\left[\frac{1}{\left(1+j\Gamma_{\mathcal{B}}\right)^2}+\frac{1}{\left(1+j\Gamma_{\mathcal{B}}+2\Gamma_{\text{asy}}\right)^2}\right. \nonumber\\ &+&\left. \frac{2}{\left(1+j\Gamma_{\mathcal{B}}+\Gamma_{\text{sym}}+\Gamma_{\text{asy}}\right)^2}\right],\label{var}
\end{eqnarray}
where $\Gamma_{X} = 2\tau_d/\tau_X$ with $X=\left\{ \text{sym},\text{asy},\mathcal{B} \right\}$, {$\tau_{\text{sym (asym)}}$ is the spin-orbit scattering time stemming from $z\rightarrow -z$ symmetric (asymmetric) SOC [\onlinecite{PhysRevLett.108.166606,PhysRevLett.120.156402,PhysRevB.99.205407,spin-valley-1,spin-valley-2}] and $\tau_{\mathcal{B}}$ is the magnetic dephasing time [\onlinecite{SM}].} Both the quantum correction to the average conductance, $\langle g_{qc} \rangle \equiv  \langle g \rangle - N/4$, and the associated  fluctuations, $\delta g=\sqrt{\text{var}[g]}$, contain crucial information on strength and symmetry of spin-orbit effects through their dependence on $\Gamma_{\text{sym}}$ and $\Gamma_{\text{asy}}$. For weak magnetic fields and weak SOC (i.e. $\tau_{\mathcal{B}},\tau_{\text{sym}},\tau_{\text{asy}} \gg \tau_{d}$), one finds a negative quantum correction ($g_{qc}\simeq-1/4$), which corresponds to standard WL. The behavior is markedly different in billiards where SOC acts as a symmetry breaking perturbation. We find several regimes, depending on the relative magnitude of spin-orbit scattering times. If the electron dwell time is much greater than the spin-flip scattering time, $\tau_{\text{asy}}$, the spin rotational invariance is broken and the quantum correction to the average conductance becomes positive; that is, the system exhibits WAL. Using Eqs.~(\ref{wl}-\ref{var}), we easily find $g_{qc}\simeq 0.125$ and UCF amplitude $\delta g \simeq 0.176$. In contrast, for $\tau_{\text{asy}}\gg \tau_d \gg \tau_{\text{sym}}$, the quantum correction to the average conductance is strongly suppressed ($\langle g_{qc} \rangle \simeq 0$) and $\delta g \simeq 0.250$, which is characteristic of quantum dots with scattering matrix distributed according to the circular unitary ensemble  [\onlinecite{RevModPhys.69.731}]. The effective breaking of time reversal symmetry in this regime can be understood from the phenomenology of SOC in honeycomb layers. The spin-valley coupling (present in materials with broken sublattice symmetry, such as group-VI dichalcogenides) acts at low energies as a valley-Zeeman field, thus mimicking a magnetic field [\onlinecite{note2}]. We compiled our results in Table \ref{tabela}.


To validate the random matrix theory predictions, we perform numerically exact real-space simulations of the magnetoconductance of the chaotic system. For simplicity, we consider a graphene nanostructure proximity coupled to a high-SOC semiconductor; see Fig.\,\ref{Imagem1}(a). The Hamiltonian of the quantum dot can be expressed as $H=H_{g}+H_{\text{SO}}$, where $H_{g}$ describes the usual nearest-neighbor hopping between $p_z$-orbitals and $H_{\text{SO}}=H_{\text{sym}}+H_{\text{asy}}$ captures the proximity-induced SOC [\onlinecite{PhysRevB.93.155104,PhysRevB.97.085413,PhysRevB.98.045407}]. (We neglect the intrinsic Kane-Mele SOC of graphene [\onlinecite{PhysRevLett.122.046403}], which is too weak to cause any significant perturbation to the quantum dot.) In terms of annihilation (creation) operators $c_{i,\sigma}$  ($c_{i,\sigma}^\dagger$) that remove (add) electrons to site $i$ with spin $\sigma = \uparrow,\downarrow$,  the terms $H_g$, $H_{\text{sym}}$ and $H_{\text{asy}}$ read as  
\begin{eqnarray}
 H_g&=&-\sum_{\langle i,j \rangle,\sigma} t^{ij} \, c_{i,\sigma}^{\dagger} c_{j,\sigma}\,, 
\\
H_{\text{sym}}&=& - \sum_{\langle\langle i,j \rangle\rangle, \sigma} \frac{\imath \lambda_{\text{sym}}^{ij}}{3\sqrt{3}} \, c_{i,\sigma}^{\dagger} \left[s_z \right]_{\sigma \sigma}  c_{j,\sigma}\,,
\label{TBH-sym}
\\
H_{\text{asy}}&=& - \sum_{\langle i,j \rangle, \sigma, \sigma^\prime } \frac{2\imath \lambda_{\text{asy}}^{ij}}{3} \, c_{i,\sigma}^{\dagger}\left(\left[\mathbf{s}\right]_{\sigma \sigma^\prime}\times \hat \mathbf{r}_{ij}\right)_z c_{j,\sigma^\prime}\,,
\label{TBHC}
\end{eqnarray}
where the indices $i$ and $j$ run over all lattice sites,\,$ \langle \cdots \rangle$ ($\langle\langle \cdots \rangle\rangle$) denotes a sum over nearest-neighbor (next-nearest-neighbor) sites, $\hat \mathbf{r}_{ij}$ is the unit vector along the line segment connecting the sites $i$ and $j$ and $t^{ij}=t e^{\imath \phi_{ij}}$, $\lambda_{\text{asy}}^{ij}=\lambda_{\text{BR}} e^{\imath \phi_{ij}}$ and $\lambda_{\text{sym}}^{ij}=\lambda_{\text{sv}} \delta_i \nu_{ij} e^{\imath \phi_{ij}}$ are Peierls’ substitution modified hopping integrals with phases $\phi_{ij}=(e/\hbar) \int_{\mathbf{r}_i}^{\mathbf{r}_j} \mathbf{A} \cdot d \mathbf{r}$. Here,   $\lambda_{\text{sv}(\text{BR})}$ is the spin-valley (BR) coupling strength and $\mathbf{A} = - B_{\perp} \hat \mathbf{y}$ is the magnetic vector potential in the Landau gauge. Furthermore, the $\nu_{ij}$ are signs that equal $\pm 1$ if the electron hops clockwise (anticlockwise) to a next-nearest site within a given hexagonal plaquette and $\delta_{i}=\pm 1$ distinguishes between the sublattices $A$($B)$ [\onlinecite{PhysRevB.95.165415,PhysRevLett.120.156402}].

The billiard is constructed by cutting a half-stadium connected to two identical leads out of a graphene sheet [\onlinecite{PhysRevLett.102.056806,PhysRevB.84.205421}]. To break the left-right symmetry, we cut out circular segments at the top left and bottom right in a way that the graphene lattices are terminated abruptly (see Fig.(\ref{Imagem1}-a)). Before attempting to confirm the predicted statistical behavior of the conductance, we verify that the simulated dots support the universal quantum transport regime, $\tau_d \gg \tau_{\text{erg}}$. To estimate the dwell time, we determine the spectrum of closed cavities with an area $\mathcal A \approx$ $1.2 \times 10^3$ nm$^2$ (containing around $10^5$ energy levels).  The calculated mean level spacing ($\Delta$) ranges from $0.2$ to $0.4$ meV, depending on the specific SOC parameter values; for additional details see the SM [\onlinecite{SM}]. This translates into dwell times ($\tau_d \approx \pi\hbar/N\Delta$) on the order of $\approx 10/N$ ps. Meanwhile, the electron transit time in the graphene dot is simply $\tau_{\text{erg}}\approx \sqrt{\mathcal A}/v$, where $v=3 a t/2\hbar \approx 10^6$ m/s (assuming a typical hopping integral $t=2.8$ eV and a lattice constant $a$ of 0.25 nm). As a result,  $\tau_d/\tau_{\text{erg}} \gg 1$ is always satisfied for typical ballistic point contacts with small number of open channels $N\approx 1-10$. 

For our numerical study, we use the recursive Green's function formalism [\onlinecite{Lewenkopf2013,PhysRevB.98.155407,PhysRevB.102.041107}] as implemented in the Kwant code [\onlinecite{Groth_2014}]. 
From the earlier theoretical analysis, the spin-orbit effects are expected to influence the statistical behavior of the conductance whenever the spin-orbit scattering time,  $\tau_{\text{so}}= (\tau_{\text{asy}}^{-1}+\tau_{\text{sym}}^{-1})^{-1}$, is short compared with the cavity dwell time. The conductance fluctuation $\delta g$  is shown in Fig.\ref{Imagem1} (b,c) for selected parameters. In the absence of SOC, the fluctuations are consistent with the circular orthogonal ensemble  ($ \delta g \simeq 0.35$) at zero field  and with the circular unitary ensemble  ($\delta g \simeq 0.25$) for a magnetic   flux $\Phi$ on the order of the quantum of flux [\onlinecite{PhysRevLett.102.056806}]. After the proximity-induced SOC is turned on, the statistical properties of the quantum dot are seen to critically depend on the relative magnitude of the spin-orbit effects in complete accord with our prediction [Eqs. (\ref{wl})-(\ref{var})]. A transition to the circular symplectic ensemble ($\delta g \simeq 0.176$) is observed for sufficiently strong BR coupling in the low-field regime ($\tau_d \gg \tau_{\mathcal B}$). This behavior is robust against the presence of symmetric-type SOC as long as the BR effect remains as a strong perturbation ($\tau_{\text{asy}} \ll \tau_d$). Moreover,\,in quantum dots with weak BR effect and strong spin-valley coupling, time reversal symmetry is effectively broken and, as a result, the conductance fluctuation approaches the circular unitary ensemble prediction at zero field ($\delta g \simeq 0.250$). Interestingly, the combined effect of a strong spin-valley coupling and a high magnetic field reduces the conductance fluctuation down to $\delta g \simeq 0.125$, which is lower than the UCF value in any of the Wigner-Dyson ensembles. The orthogonal to unitary/symplectic ensemble transitions facilitated by SOC and the associated statistical properties are summarized in Table \ref{tabela}. It is important to note that due to the small quantum dot size in the simulations, the ensemble transitions are observed at rather large SOC values ($\approx$ 0.1 eV). On the other hand, the experimentally achievable proximity-induced SOC energy scales are more modest (in graphene on a group-VI dichalcogenide monolayer, these range from 0.1 to 10 meV depending on the high-SOC material used and the quality of the interface  [\onlinecite{10.1038/ncomms9339, PhysRevB.93.155104,PhysRevB.95.165415,PhysRevB.98.045407}]). Hence, the experimental validation of our findings would require dots of substantially larger dimensions to ensure  $\tau_{\text{so}} \ll \tau_{d}$. 

\begin{figure}[!]
\centering
\includegraphics[width=0.8\linewidth]{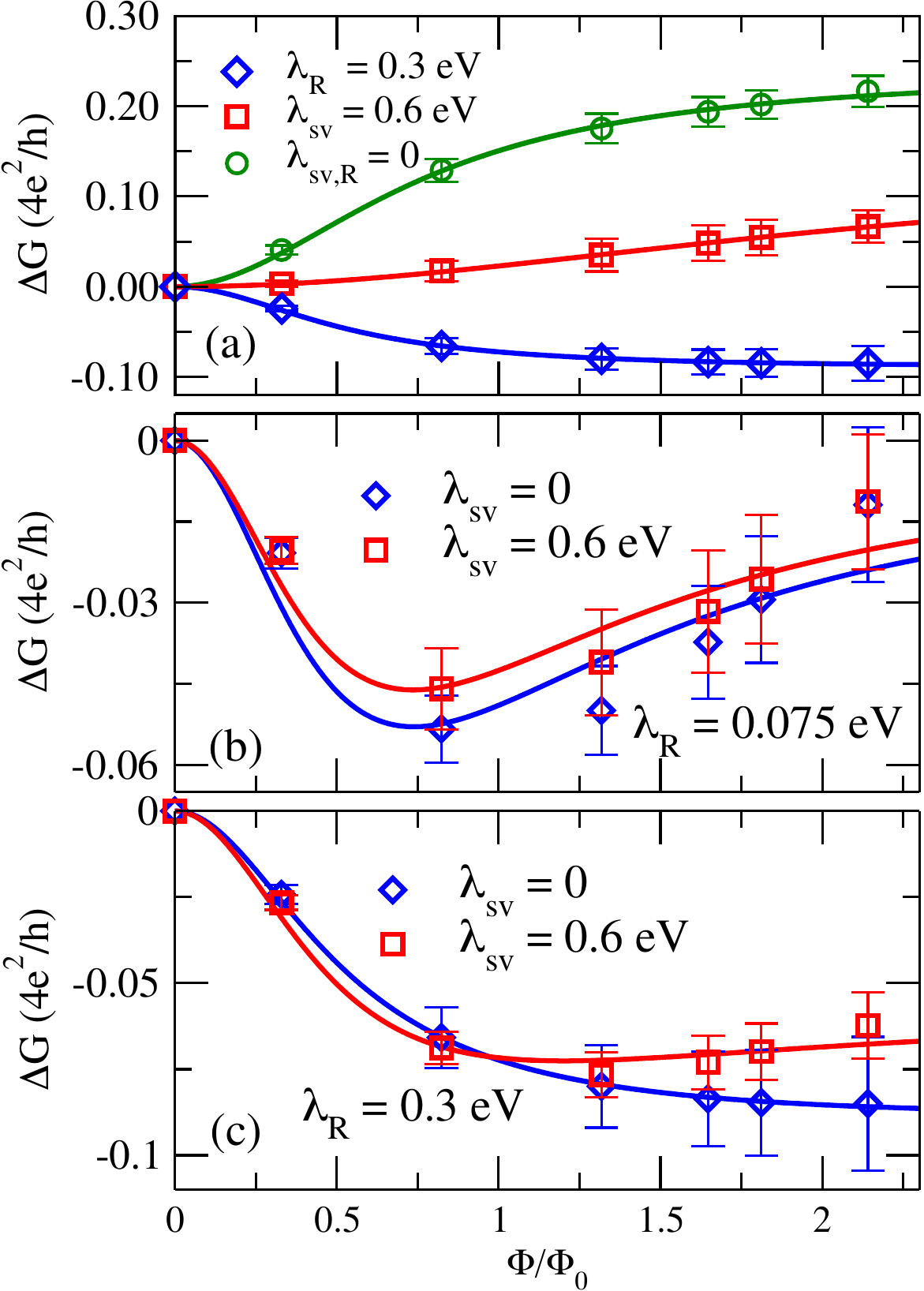}
\caption{(a-c) Average magnetoconductance as a function of the magnetic flux obtained numerically for selected SOC parameters. Solid lines are fits to Eq.(\ref{wl}) [\onlinecite{SM}]. Data points are calculated using 20 chaotic samples averaged over the Fermi energy window $[0.45,1.50]$ eV. Here $\Phi_0=h/e$ is the quantum of flux.}
\label{Imagem3}
\end{figure}

Now we turn to the magnetotransport fingerprints of proximity-induced SOC. In Fig. (\ref{Imagem3}), we show the calculated magnetoconductance $\Delta G (B_{\perp}) = \langle G (B_{\perp}) - G (0) \rangle$. The negative signal (WAL) observed in  all simulated quantum dots with sizable BR coupling provides a clear signature of the   orthogonal-to-symplectic ensemble transition. We note that this is a direct numerical evaluation of conductance WAL corrections in a chaotic billiard. In accord with the theory [c.f. Eq.\,(\ref{wl})], the BR-coupling-induced negative quantum correction is seen to be robust to the presence of symmetric SOC; see Figs. \ref{Imagem3} (b)-(c). The special role played by the spin-valley coupling in quantum dots with weak or vanishing BR effect ($\tau_{\text{asy}}\gg\tau_d$) is also borne out by the simulations. Indeed, the simulation with $\lambda_{\text{BR}}=0$ and $\lambda_{\text{sv}}=0.6$ eV shows a clear suppression of quantum interference effects due to spin-valley coupling since, as discussed earlier, the latter acts on the ballistic electrons as a valley-Zeeman field. The underlying spin-orbit scattering times can be estimated by fitting the numerical data to Eq.(\ref{wl}). We find $\tau_{\text{so}}$ to be on the order of 1 ps, which puts the simulated devices within the universal regime where the theory is expected to be accurate; see the SM for additional details [\onlinecite{SM}].
We note in passing that in the absence of SOC, the magnetoconductance can be accurately fitted to the well-known expression $\Delta G / (4e^2/h)=  \mathcal{G}(1+\frac{\tau_{\mathcal{B}}}{2\tau_d})^{-1}$ [\onlinecite{PhysRevLett.70.3876}] with $\mathcal{G}=0.23$, in excellent agreement with the random matrix theory prediction ($\mathcal{G}_{\text{RMT}}=0.22$).

To put our predictions into context, we first note that diffusive WAL behavior in (non-chaotic) graphene devices with interface-induced SOC is now well established [\onlinecite{PhysRevX.6.041020,Yang_2016,Yang2017,Volkl2017,Zihlmann2018,Wakamura2018,PhysRevLett.108.166606, PhysRevB.99.205407}]. Transition metal dichalcogenides represent a broad family of high-SOC layered materials, which can be used to fabricate the envisaged chaotic Dirac-Rashba billiard characterized by competing spin-orbit effects with different symmetries. According to our findings, chaotic billiards built from graphene-based heterostructures can display robust signatures of WAL in the universal regime of quantum transport provided that the asymmetric spin-orbit scattering time is shorter than the dwell time of the cavity. We expect that such a condition can be achieved by fabricating mesoscopic quantum dots with linear size approaching the typical (bulk) mean free paths. Electronic transport measurements on submicrometer graphene quantum dots  
have been recently reported [\onlinecite{GQD-exp1,GQD-exp2,GQD-exp3,GQD-review}], which gives us extra confidence that the predictions in this paper can be put to the test in the near future.

In summary, we have used random matrix theory to investigate the statistical behavior of the average conductance and its universal fluctuations in chaotic graphene-based billiards with proximity-induced SOC. Our study, supported by real-space quantum transport calculations, shows that the proximity-induced SOC strongly influences the device conductance in zero and finite applied magnetic fields. Quantum dots with a sizable BR effect (i.e. with asymmetric spin-orbit scattering time shorter than the cavity dwell time) were found to display robust WAL signals with fluctuations consistent with the circular-symplectic ensemble.

\begin{acknowledgments}
A.L.R.B. and J.G.G.S.R. were supported by CNPq (Grant No. 307474/2018-6) and FACEPE (Grant No. APQ-0325-1.05/18). A.F. gratefully acknowledges the financial support from the Royal Society through a Royal Society University Research Fellowship. 
\end{acknowledgments}

\bibliography{ref}

\begin{thebibliography}{71}
\bibitem[{not({\natexlab{a}})}]{note1}
\bibinfo{note}{Additional SOCs are possible if low-symmetry transition metal
  dichalcogenides are used [\onlinecite{PhysRevB.98.045407}].}
\bibitem[{not({\natexlab{b}})}]{note2}
\bibinfo{note}{This was first noted by McCann and Fal'ko within the context of
  diffusive (non-chaotic) quantum transport
  [\onlinecite{PhysRevLett.108.166606}].}
\end{thebibliography}

\end{document}


\title{Supplemental Material: Effect of proximity-induced spin-orbit coupling in graphene mesoscopic billiards}

\author{Anderson L. R. Barbosa$^1$, Jorge Gabriel G. S. Ramos$^2$, Aires Ferreira$^3$}

\affiliation{$^1$Departamento de F\'{\i}sica,Universidade Federal Rural de Pernambuco, 52171-900 Recife, Pernambuco, Brazil\\$^2$Departamento de F\'isica, Universidade Federal da Para\'iba, 58051-970 Joa\~ao Pessoa, Para\'iba, Brazil\\
$^3$Department of Physics and York Centre for Quantum Technologies, University of York, York YO10 5DD, United Kingdom}






\maketitle

\section{Stub Model}

\begin{figure}[!]
\includegraphics[width=0.4\linewidth]{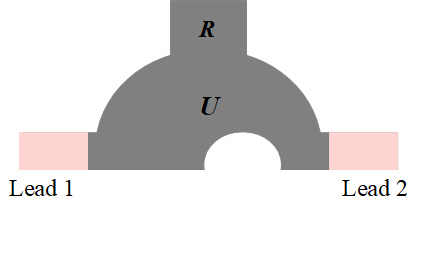}
\includegraphics[width=0.7\linewidth]{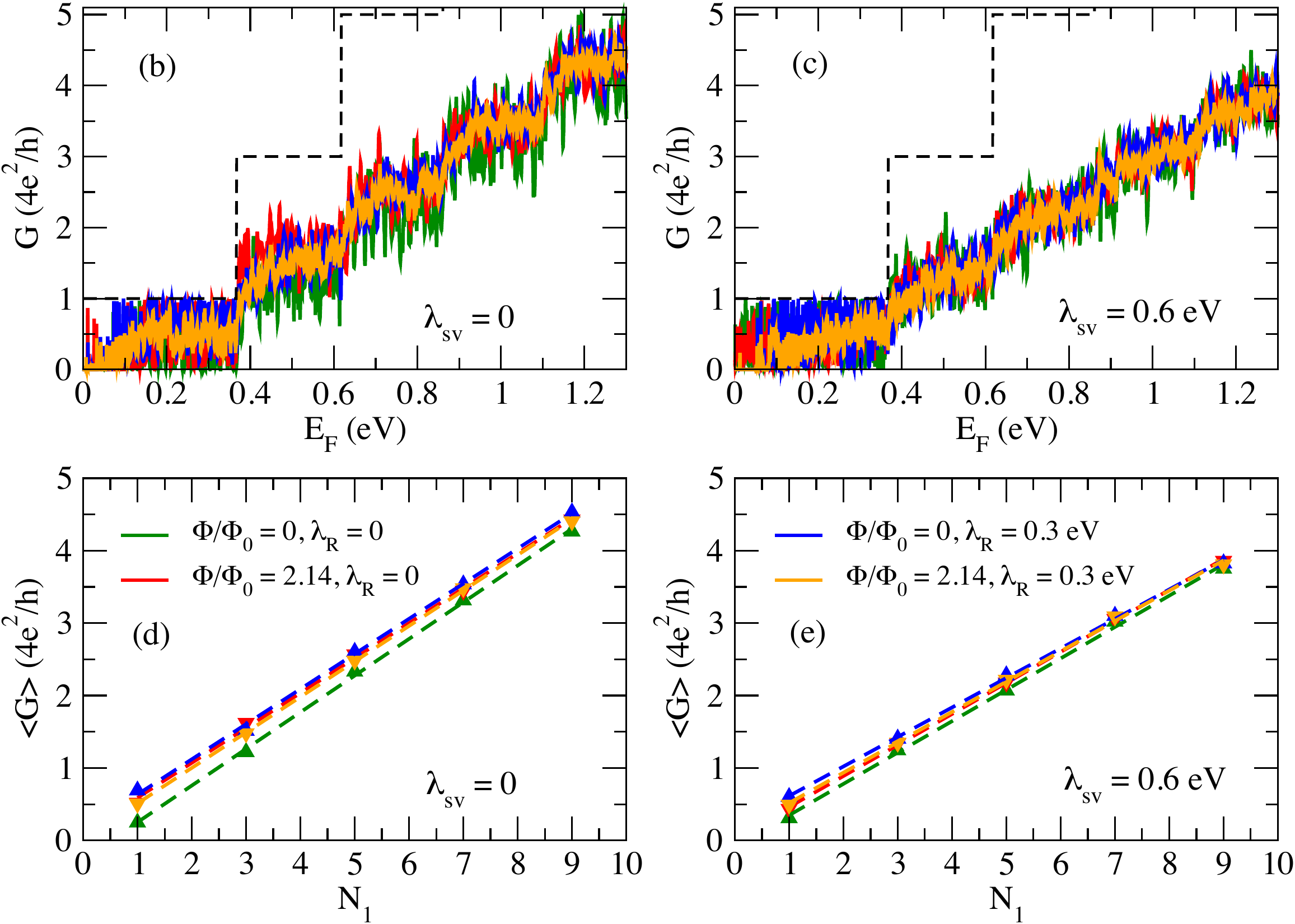}
\caption{ (a) Schematic illustration of quantum dot (represented by $U$ matrix) connected to stub (represented by $R$ matrix) and two leads.  (b-c) Conductance as a function of Fermi energy at selected SOC and magnetic field values. Dashed lines indicate number of propagating wave channels in the leads, $N_1$. (d-e) Average conductance as a function of number of open channels. Data points are calculated using 20 chaotic billiard realizations.}
\label{Imagem1}
\end{figure}

The stub model has been applied to quantum dots of GaAs [\onlinecite{PhysRevB.68.125329,PhysRevB.65.081302,PhysRevB.84.035453}] and pristine graphene  [\onlinecite{PhysRevB.93.125136,PhysRevB.99.195131}]. Here, we extend the formulation to graphene with pseudo-spin--spin coupling due to SOC. In the stub model, the scattering matrix of an isolated quantum dot with $M$ energy levels   is parameterized as follows [\onlinecite{RevModPhys.69.731}]
\begin{eqnarray}
\mathcal{S} = PU(1 - Q^{\dagger}RQU)^{-1}P^{\dagger},
\end{eqnarray}
where $U$ is a $4M \times 4M$ random unitary symmetric matrix taken from the
circular orthogonal ensemble. The matrix $R$ is an unitary matrix with dimension $4(M-N)$, which is used to introduce external perturbations, such as a magnetic field and SOC, via the "stub" [\onlinecite{RevModPhys.69.731}]; see Fig. \ref{Imagem1} (a). The $P$ and $Q$ are $4N \times 4M$ and  $(4M-4N) \times 4M$ projection matrices defined as $P_{ij} = \delta_{i,j}$ and $Q_{ij} = \delta_{i+4N,j}$ [\onlinecite{PhysRevB.65.081302}]. These matrices connect the leads ($P$) and stub ($Q$) to the quantum dot. Hence, the scattering matrix $\mathcal{S}$ has dimension $4N \times 4N$.
The matrix $R$ is given by [\onlinecite{PhysRevB.65.081302}]
\begin{eqnarray}
 R(\varepsilon,\vec \mathcal{B}) =  \exp\left[\frac{2 \pi \imath}{M\Delta} 
  \left(\varepsilon - H'(\vec \mathcal{B})\right)\right],\label{R}
\end{eqnarray}
where $\Delta$ is the mean level spacing of the quantum dot and $H'$ is an $(4M-4N)$ dimensional matrix encoding the perturbations to the bare Hamiltonian [\onlinecite{PhysRevB.93.125136}], which for a proximitized graphene quantum dot with SOC is given by
\begin{eqnarray}
H'(\vec \mathcal{B})&=& \frac{\sqrt{N}\Delta}{2\pi}\left[\frac{\imath}{2} \sqrt{\frac{\tau_d}{\tau_{\mathcal{B}}}} \left(A_1 \sigma_x + A_2 \sigma_y\right) \otimes s_0 
+ \imath \sqrt{\frac{\tau_d}{\tau_{\text{sym,I}}}} I_1 \sigma_z \otimes s_z \right.\nonumber\\&+& \left. \imath \sqrt{\frac{\tau_d}{\tau_{\text{sym,sv}}}} I_2 \sigma_0 \otimes s_z + i \sqrt{\frac{\tau_d}{\tau_{\text{asy}}}}  \left(B_1 \sigma_x \otimes s_y +B_2\sigma_y \otimes s_x \right)\right].\label{HRMT}
\end{eqnarray}
Here, $s_i (\sigma_i)_{i=x,y,z}$ are Pauli matrices acting on the spin (pseudospin) spaces and $\tau_{\text{sym,vz}}$, $\tau_{\text{sym,I}}$ and $\tau_{\text{asy}}$ are characteristic scattering times associated with spin-valley coupling, Kane-Mele SOC and $z\rightarrow -z$ asymmetric BR interaction, respectively. Moreover, $A_i$, $I_i$ and $B_i$ ($i=1,2$) are real antisymmetric matrices of dimension ($(M-N) \times (M-N)$) that satisfy $\langle \textbf{Tr}\left(A_iA_j^T\right) \rangle = \langle \textbf{Tr}\left(I_{i}I_j^T\right) \rangle = \langle \textbf{Tr}\left(B_{i}B_j^T\right) \rangle = \delta_{ij} M^2$. The mean dwell time is given by $\tau_d = 2\pi \hbar/N\Delta$, while the magnetic dephasing rate is defined as $\tau_{\mathcal{B}}^{-1}=\tau_{e}^{-1}\times 2\pi c \left(\Phi/\Phi_0\right)^2$ [\onlinecite{PhysRevB.84.205421}] with $c$ a system-dependent parameter of order unity, while $\Phi$ denotes the magnetic flux (here, $\Phi_0=h/e$ is the quantum of flux). 

The average of the conductance, Eq.(2) of main text, can be obtained using the relation
\begin{eqnarray}
\left< \mathcal{S}_{ij;\alpha \beta}(\epsilon,\vec \mathcal{B}) \, \mathcal{S}_{i'j';\alpha' \beta'}^*(\epsilon,\vec \mathcal{B})\right> = \delta_{ii'}\delta_{jj'}\mathcal{D}_{\alpha \beta; \beta'\alpha'}+ \delta_{ij'}\delta_{ji'}\left(\mathcal{T}\mathcal{C}\mathcal{T}\right)_{\alpha \beta; \alpha'\beta'}\label{SS}
\end{eqnarray}
which is valid in the semiclassical limit, $N\gg 1$. The matrix $\mathcal{T}$ is defined as $\mathcal{T} = \sigma_0 \otimes s_0 \otimes \sigma_0 \otimes s_y$, 
while the matrices $\mathcal{D}$ and $\mathcal{C}$ encode contributions of Diffuson and Cooperon diagrams, respectively, given by
\begin{eqnarray}
\mathcal{D}^{-1}&=&M\sigma_0 \otimes s_0 \otimes \sigma_0 \otimes s_0 - \textbf{Tr}\left(R\otimes R^\dagger \right),\nonumber\\
\mathcal{C}^{-1}&=&M\sigma_0 \otimes s_0 \otimes \sigma_0 \otimes s_0 - \textbf{Tr}\left(R\otimes R^\star \right),\label{DC}
\end{eqnarray}
where $\dagger$ ($\star$) denotes Hermitian (complex) conjugation. The evaluation of the trace operations in Eq. (\ref{DC}) is carried out using the identity
$$\left(\sigma_i \otimes s_j \otimes \sigma_k \otimes s_l\right)  \left(\sigma_i' \otimes s_j' \otimes \sigma_k' \otimes s_l'\right)=\left(\sigma_i\sigma_i'\right) \otimes \left(s_j' s_j \right) \otimes \left(\sigma_k' \sigma_k \right) \otimes \left(s_l s_l'\right).$$
From Eqs. (\ref{R}) -(\ref{HRMT}), we obtain, in the limit $M\gg N$, 
\begin{eqnarray}
\mathcal{D}^{-1}&=&N\left[\left(1+\frac{\tau_d}{\tau_{\mathcal{B}}}+\frac{\tau_d}{\tau_{\text{sym,I}}}+\frac{\tau_d}{\tau_{\text{sym,sv}}}+\frac{2\tau_d}{\tau_{\text{asy}}}\right)\sigma_0 \otimes s_0 \otimes \sigma_0 \otimes s_0 \right.\nonumber\\
&-& \left.\frac{1}{2}\frac{\tau_d}{\tau_{\mathcal{B}}} \left(\sigma_x \otimes s_0 \otimes \sigma_x \otimes s_0 +\sigma_y \otimes s_0 \otimes \sigma_y \otimes s_0  \right) - \frac{\tau_d}{\tau_{\text{sym,I}}} \sigma_z \otimes s_z \otimes \sigma_z \otimes s_z \right.\nonumber\\
&-& \left. \frac{\tau_d}{\tau_{\text{sym,sv}}} \sigma_0 \otimes s_z \otimes \sigma_0 \otimes s_z - \frac{\tau_d}{\tau_{\text{asy}}}\left(\sigma_x \otimes s_y \otimes \sigma_x \otimes s_y +\sigma_y \otimes s_x \otimes \sigma_y \otimes s_x  \right)\right]\label{Diffuson}\\
\mathcal{C}^{-1}&=&N\left[\left(1+\frac{\tau_d}{\tau_{\mathcal{B}}}+\frac{\tau_d}{\tau_{\text{sym,I}}}+\frac{\tau_d}{\tau_{\text{sym,sv}}}+\frac{2\tau_d}{\tau_{\text{asy}}}\right)\sigma_0 \otimes s_0 \otimes \sigma_0 \otimes s_0 \right.\nonumber\\
&+& \left.\frac{1}{2}\frac{\tau_d}{\tau_{\mathcal{B}}} \left(\sigma_x \otimes s_0 \otimes \sigma_x \otimes s_0 +\sigma_y \otimes s_0 \otimes \sigma_y \otimes s_0  \right) - \frac{\tau_d}{\tau_{\text{sym,I}}} \sigma_z \otimes s_z \otimes \sigma_z \otimes s_z \right.\nonumber\\
&-& \left. \frac{\tau_d}{\tau_{\text{sym,sv}}} \sigma_0 \otimes s_z \otimes \sigma_0 \otimes s_z - \frac{\tau_d}{\tau_{\text{asy}}}\left(\sigma_x \otimes s_y \otimes \sigma_x \otimes s_y +\sigma_y \otimes s_x \otimes \sigma_y \otimes s_x  \right)\right].
\label{Cooperon}
\end{eqnarray}

Replacing Eqs.(\ref{Diffuson}-\ref{Cooperon}) into Eq. (\ref{SS}), we easily obtain
\begin{eqnarray}
\langle G \rangle =  \frac{4e^2}{h}\frac{N_1N_2}{N} +  \frac{2e^2}{h} \frac{N_1N_2}{N^2} \left[\frac{1}{1+\Gamma_{\mathcal{B}}}-\frac{1}{1+\Gamma_{\mathcal{B}}+2\Gamma_{\text{asy}}}- \frac{2}{1+\Gamma_{\mathcal{B}}+\Gamma_{\text{asy}}+\Gamma_{\text{sym}}}\right]  \label{G}
\end{eqnarray}
and
\begin{eqnarray}
 \text{var} [G] &=&  \frac{4e^4}{h^2} \frac{N_1^2N_2^2}{N^4} \left[1+\frac{1}{\left(1+2\Gamma_{\text{asy}}\right)^2}+ \frac{2}{\left(1+\Gamma_{\text{sym}}+\Gamma_{\text{asy}}\right)^2}\right.\nonumber\\
&+&\left.\frac{1}{\left(1+\Gamma_{\mathcal{B}}\right)^2}+\frac{1}{\left(1+\Gamma_{\mathcal{B}}+2\Gamma_{\text{asy}}\right)^2}+ \frac{2}{\left(1+\Gamma_{\mathcal{B}}+\Gamma_{\text{sym}}+\Gamma_{\text{asy}}\right)^2}\right].\label{varG}
\end{eqnarray}
Equations (\ref{G}) and (\ref{varG}) are equivalent to Eqs. (4) and (5) in the main text, where $N_1=N_2=N/2$.

\section{Numerically exact results}

Here, we present additional details on the real-space simulations performed for the ballistic chaotic system, Fig.(\ref{Imagem1}). Figure ~\ref{Imagem1} shows the conductance as a function of the Fermi energy ((b)-(c)) and as a function of open channels ((c)-(d)) for selected  parameters. From the fit to the data in the Fig.(\ref{Imagem1}-d,e) (dashed lines), we found $ \langle G \rangle/4e^2/h \approx 0.5 \times N$ for $\Phi/\Phi_0 = 2.14$, and $ \langle G \rangle/4e^2/h \approx 0.5 \times N +  \langle G_{\text{qc}} \rangle $, in the absence of magnetic flux, with $ G_{\text{qc}}$ signaling weak-localization ($ G_{\text{qc}}>0$) and anti-weak-localization ($ G_{\text{qc}}<0$) for $\lambda_{\text{R}}=0$ and $\lambda_{\text{R}}=0.3$ eV, respectively.

\begin{table*}
\centering
\begin{tabular}{ccccccccccc}
\hline
    \hline
Fig. (2) \quad\quad &  $\lambda_{\text{R}}$ (eV)\quad\quad & $\lambda_{\text{sv}}$ (eV)\quad\quad & $\Delta$ (meV) \quad\quad & $\tau_d$ (ps) \quad\quad &$\tau_{\text{asy}}$ (ps) \quad\quad &$\tau_{\text{sym,sv}}$ (ps) \quad\quad & $\tau_{\text{so}}$ (ps) & Limit & Ensemble\\\hline
    \hline
    \hline
(a) & 0.0 & 0.0 & 0.43 & 0.69 & $\times$ & $\times$& $\times$&$\tau_{\text{so}}\gg \tau_d$& COE\\\hline
(a) & 0.0 & 0.6 & 0.22 & 1.33 & $\times$ & $0.16$& $0.16$ & $\tau_d \gg \tau_{\text{so}}$& CUE\\\hline
(a) & 0.3 & 0.0 & 0.21 & 1.44 & $0.08$ & $\times$& $0.08$ & $\tau_d \gg \tau_{\text{so}}$& CSE\\\hline\hline
(b) & 0.075 & 0.0 & 0.22 & 1.35 & $1.60$ & $\times$& $1.60$& $\tau_d \approx \tau_{\text{so}}$& \\\hline
(b) & 0.075 & 0.6 & 0.22 & 1.34 & $2.18$ & $3.95$& $1.41$& $\tau_d \approx \tau_{\text{so}}$& \\\hline\hline
(c) & 0.3 & 0.0 &  0.21 & 1.44 & $0.08$ & $\times$& $0.08$& $\tau_d \gg \tau_{\text{so}}$& CSE\\\hline
(c) & 0.3 & 0.6 & 0.20 & 1.44 & $1.26$ & $1.26$& $0.63$& $\tau_d \gg \tau_{\text{so}}$& CSE\\\hline
    \hline\hline
\end{tabular}
\caption{BR ($\lambda_{\text{R}}$) and  spin-valley ($\lambda_{\text{sv}}$) SOC parameters used in the numerical simulation reported in Fig. 2, main text. The  calculated mean space level $\Delta$ within the Fermi energy window $[-0.5t,0.5t]$ and the relevant time scales estimated from the fit to the magnetoconductance data are also shown. }\label{tabela2}
\end{table*}

\bibliography{ref}